\documentclass[twocolumn,showkeys,a4paper,10pt]{revtex4}
\usepackage{amsthm}
\usepackage{amsmath}
\usepackage{mathrsfs}
\usepackage{graphicx}% Include figure files
\usepackage{dcolumn}% Align table columns on decimal point
\usepackage{bm}% bold math
\usepackage{subfigure}
\usepackage{setspace}  
\begin{document}

\markboth{Wen-An Li}
{Enhancement of optomechanically induced sum sideband using parametric interactions}

\linespread{1.6}

\title{Enhancement of optomechanically induced sum sideband using parametric interactions}

\author{Wen-An Li$^1$\footnote{E-mail: liwenan@126.com, liwa@gzhu.edu.cn} and Guang-Yao Huang$^2$
} 
\address{1.School of Physics and Electronic Engineering, Guangzhou University, Guangzhou 510006, China\\
2.Institute for Quantum Information \& State Key Laboratory of High Performance Computing, College of Computer, National University of Defense Technology, Changsha 410073, China}

\begin{abstract}
\begin{spacing}{1}
We theoretically study radiation pressure induced generation of the frequency components at the sum sideband in an optomechanical system containing an optical parametric amplifier (OPA). It is shown that an OPA inside a cavity can considerably enhance the amplitude of sum sideband even with low power input fields. We find a new matching condition for the upper sum sideband generation. The height and width of the new peak can be adjusted by the nonlinear gain of the OPA. Furthermore, the lower sum sideband generation can be enhanced with several orders of magnitude by tuning the nonlinear gain parameter and the phase of the field pumping the OPA. The enhanced sum sideband may have potential applications to the manipulation of light in a on-chip optomechanical device and the sensitively sensing for precision measurement in the weak optomechanical coupling regime. 
\end{spacing}
\end{abstract}
\maketitle
\section{introduction}
In recent years, optomechanical systems~\cite{1,2,3} have received considerable theoretical~\cite{4,5,6,7,8,9} and experimental~\cite{10,11,12,13} interest due to the potential applications to the study of a range of topics such as gravitational wave detection~\cite{14,15,16}, tiny displacement measurement~\cite{17,18} and cooling of mechanical oscillators~\cite{19,20,21,22,23}. Among these applications, optomechanically induced transparency (OMIT)~\cite{10,24,25,26} is a very interesting phenomenon, which is an analog of electromagnetically induced transparency. OMIT is a kind of induced transparency caused by radiation pressure coupling of an optical and a mechanical mode, where the anti-Stokes scattering of an intense red-detuned optical control field brings about a modification in the optical response of the optomechanical cavity making it transparent in a narrow bandwidth around the cavity resonance for a probe beam. OMIT can be explained by the Heisenberg-Langevin equations, which are nonlinear and very hard to get an analytic solution. If the probe field is much weaker than the control field, one can use the perturbation method to get the prominent feature of  optomechanically induced transparency. Recently, nonlinear optomechanical dynamics have emerged as an interesting frontier in cavity optomechanics~\cite{27,28,29,30}. This emerging subject leads to a variety of high-order OMIT effects due to the intrinsic nonlinear optomechanical interactions~\cite{31,32}, such as photon-phonon polariton pairs~\cite{33} and sideband generations~\cite{27,34,35}. In particular, OMIT with multiple probe fields driven has also been explored~\cite{36,37}. Generation of spectral components at sum (or difference)  sideband are demonstrated analytically, which may have great potential in the precise measurement of parameters and phonon number of optomechanical systems~\cite{38,39,40,41}. However, the sum sidebands are generally much weaker than the probe signal and thus hard to be detected or utilized. In view of the potential applications of sum sideband generation, an interesting question is whether one can easily amplify the sum sideband generation with current experimental system parameters.

Very recently, the effects of an optical parametric amplifier (OPA) inside the cavity on the optomechanical coupling~\cite{9,42}, the normal mode splitting~\cite{43}, the cooling of the mechanical mirror~\cite{44} and the realization of strong mechanical squeezing~\cite{45} have been discussed, which showed that increasing gain of the OPA enhances the coupling between the movable mirror and the cavity field. Furthermore, the effects of OPA on multipartite entanglement~\cite{46}, force sensing for a free particle~\cite{47}, and cooling in nonlinear optomechanical systems~\cite{48} have been also discussed.

In present work, we consider the effect of OPA on the sum sideband generation in a optomechanical system which is coherently driven by a trichromatic input field consisting of a control field and two probe fields. It is shown that in the presence of OPA, compared to that in a linear resonator, the amplitude of sum sideband can be significantly enhanced even with low power input fields. Interestingly, we find a new matching condition for the upper sum sideband generation. With the increasing nonlinear gain of the OPA, the linewidth of the sum sideband window is broadened and the peak value gets larger. To explain the physical interpretation of this new matching condition, features of the mechanical oscillation at sum sideband are also discussed. Furthermore, the lower sum sideband generation can be enhanced with several orders of magnitude by tuning the nonlinear gain parameter and the phase of the field pumping the OPA. The enhanced sum sideband may be useful for the optical information processing and provides an effective way to manipulate light in a solid state architecture. The present proposal can also be applied to enhance the difference sideband generation.

The paper is organized as follows. In Sec. II, we present the theoretical description of a hybrid optomechanical system and give the derivation of the Heisenberg-Langevin equation of motion in the presence of the OPA. In Sec. III, we discuss the effect of the OPA on the sum sideband generation and analyze the results deeply. Finally, a conclusion is summarized in Sec. IV.

\begin{figure}[t]
\begin{center}
\includegraphics[width=0.4\textwidth]{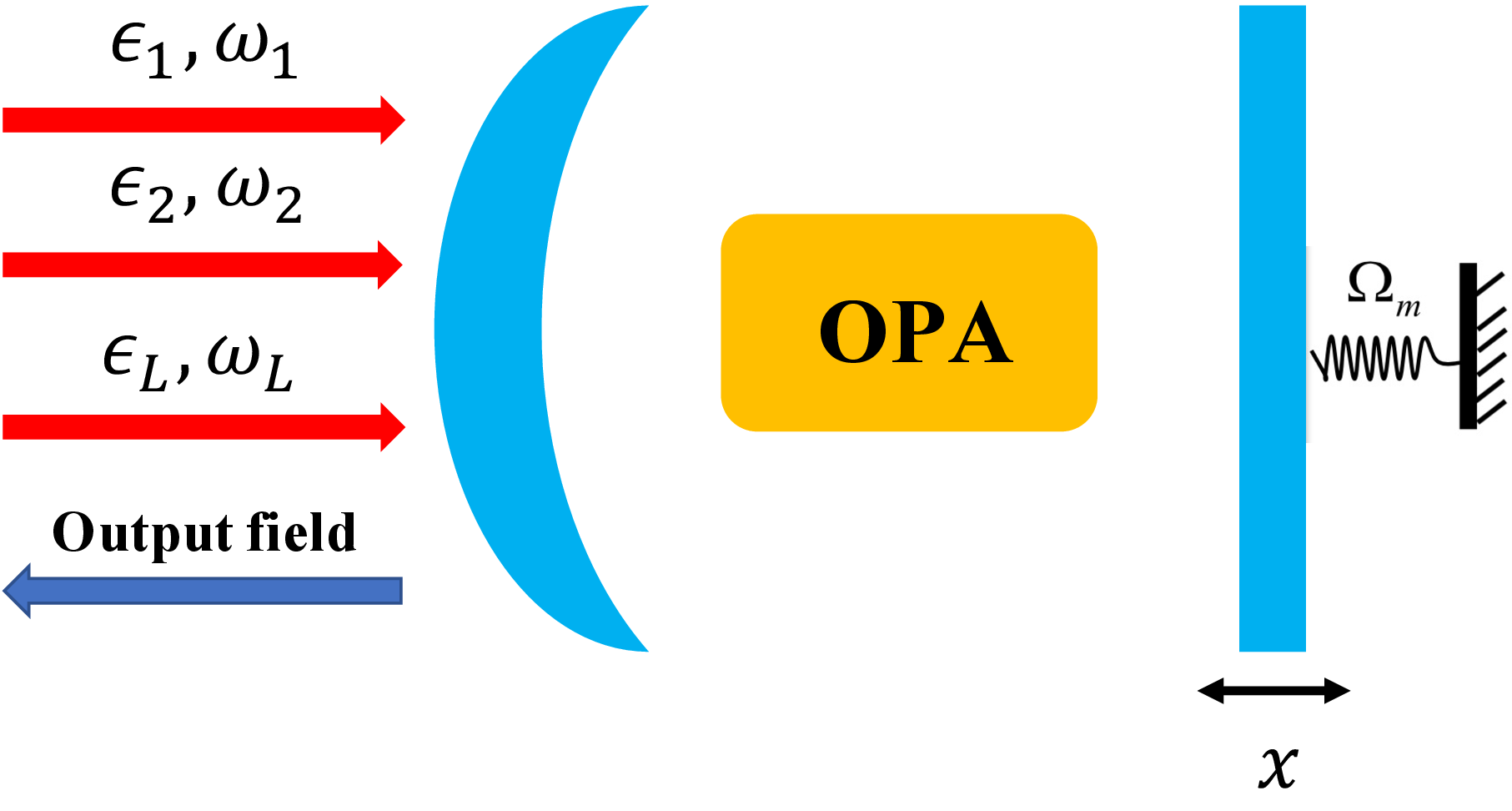}
\end{center}
\caption{Schematic diagram of an optomechanical system containing an optical parametric amplifier. The system is driven by a strong control field of frequence $\omega_L$ and two relatively weak probe laser of frequencies $\omega_1$ and $\omega_2$, respectively.}
\end{figure}
\section{theoretical model}
The model we consider is an optomechanical cavity with a vibrating mirror, which contains an optical parametric amplifier, as shown in Fig.1. The movable mirror is free to move along the cavity axis and is treated as a quantum mechanical harmonic oscillator with effective mass $m$, frequency 􏱰$\Omega_m$, and energy decay rate $\Gamma_m$. A strong pump laser of frequency $\omega_L$ and two relatively weak probe laser of frequencies $\omega_1$ and $\omega_2$ are applied to the system. The Hamiltonian formulation of such a optomechanical system reads 
\begin{align}
&\hat{H}=\hat{H}_{\mathrm{mech}}+\hat{H}_{\mathrm{opt}}+\hat{H}_{\mathrm{OPA}}+\hat{H}_{\mathrm{control}}+\hat{H}_{\mathrm{probe}},\\ \nonumber
&\hat{H}_{\mathrm{mech}}=\frac{\hat{p}^2}{2m}+\frac{1}{2}m\Omega_m^2\hat{x}^2,\\ \nonumber
&\hat{H}_{\mathrm{opt}}=\hbar\omega_c \hat{a}^\dag \hat{a}+\hbar G\hat{a}^\dag \hat{a} \hat{x},\\ \nonumber
&\hat{H}_{\mathrm{OPA}}=i\hbar g(e^{i\theta}\hat{a}^{\dag 2}e^{-2i\omega_L t}-e^{-i\theta}\hat{a}^{2}e^{2i\omega_L t}),\\ \nonumber
&\hat{H}_{\mathrm{control}}=i\hbar \sqrt{\eta_c \kappa}\epsilon_L(\hat{a}^\dag e^{-i\omega_L t}-\hat{a}e^{i\omega_L t}),\\ \nonumber
&\hat{H}_{\mathrm{probe}}=i\hbar \sqrt{\eta_c \kappa}(\epsilon_1\hat{a}^\dag e^{-i\omega_1 t}+\epsilon_2\hat{a}^\dag e^{-i\omega_2 t}-\mathrm{h.c.}),
\end{align}
where $\hat{p}$ and $\hat{x}$ describe the momentum and position of the mechanical mode, $\hat{a}$ ($\hat{a}^\dag$) is the annihilation (creation) operator of the cavity field with resonance frequency $\omega_c$. The term $\hbar G\hat{a}^\dag \hat{a} \hat{x}$ denotes the interaction between the cavity field and the movable mirror, $G$ is the the optomechanical coupling constant. $\hat{H}_{\mathrm{OPA}}$ describes the coupling of the intracavity field with OPA, $g$ is the nonlinear gain of the OPA, which is proportional to the pump power driving amplitude, $\theta$ is the phase of the field driving the OPA. $\hat{H}_{\mathrm{control}}$ and $\hat{H}_{\mathrm{probe}}$ denote the driving fields with the field amplitudes $\epsilon_k=\sqrt{P_k/\hbar \omega_k},(k=L,1,2)$, where $P_L$ is the pump power and $P_{1(2)}$ is the power of the probe field. $\kappa$ is the total loss rate which contains an intrinsic loss rate $\kappa_0$ and an external loss rate $\kappa_{\mathrm{ex}}$. The coupling parameter $\eta_c=\kappa_{\mathrm{ex}}/(\kappa_0+\kappa_{\mathrm{ex}})$ which can be continuously adjusted, is chosen to be the critical coupling $1/2$ here.

The dynamics of the system is described by a set of nonlinear Langevin equations. Since we are interested in the mean response of the system to the probe field, we write the Langevin equations for the mean values. In a frame rotating at $\omega_L$ with $\Delta=\omega_L-\omega_c$, $\delta_1=\omega_1-\omega_L$, $\delta_2=\omega_2-\omega_L$, the Heisenberg-Langevin equations can be obtain as follows:
\begin{subequations}\label{eq7}
\begin{align}
\nonumber
&\dot{a}=\left(-\frac{\kappa}{2}+i\Delta\right)a-iGxa+2ge^{i\theta}a^*\\
&+\sqrt{\eta_c \kappa}(\epsilon_L+\epsilon_1e^{-i\delta_1 t}+\epsilon_2e^{-i\delta_2 t}),\\
&m\left(\frac{d}{dt^2}+\Gamma_m\frac{d}{dt}+\Omega_m^2\right)x=-\hbar Ga^* a,
\end{align}
\end{subequations}
where the operators are reduced to their expectation values, viz, $a(t)\equiv \langle\hat{a}(t)\rangle$ and $x(t)\equiv \langle\hat{x}(t)\rangle$, the mean-field approximation by factorizing averages is used and the quantum noise terms are dropped.

Under the assumption that the input coupling laser field is much stronger than the probe field ($\epsilon_L\gg\epsilon_1, \epsilon_2$), we can use the perturbation method to deal with Eq.~(\ref{eq7}). The control field provides a steady-state solution ($\bar{a}, \bar{x}$) of the system, while the probe field is treated as the perturbation of the steady state. The total solution of the intracavity field and the mechanical displacement under both the control field and probe field can be written as $a=\bar{a}+\delta a$ and $x=\bar{x}+\delta x$. The steady state solutions of Eq.~(\ref{eq7}) can be obtained as
\begin{equation}\label{eq8}
\bar{a}=\frac{\kappa/2+i\Delta^\prime+2ge^{i\theta}}{\kappa^2/4+\Delta^{\prime 2}-4g^2}\sqrt{\eta_c\kappa}\epsilon_L,\quad \bar{x}=\frac{-\hbar G|\bar{a}|^2}{m\Omega_m^2},
\end{equation}
\begin{figure}[t]
\begin{center}
\includegraphics[width=0.28\textwidth]{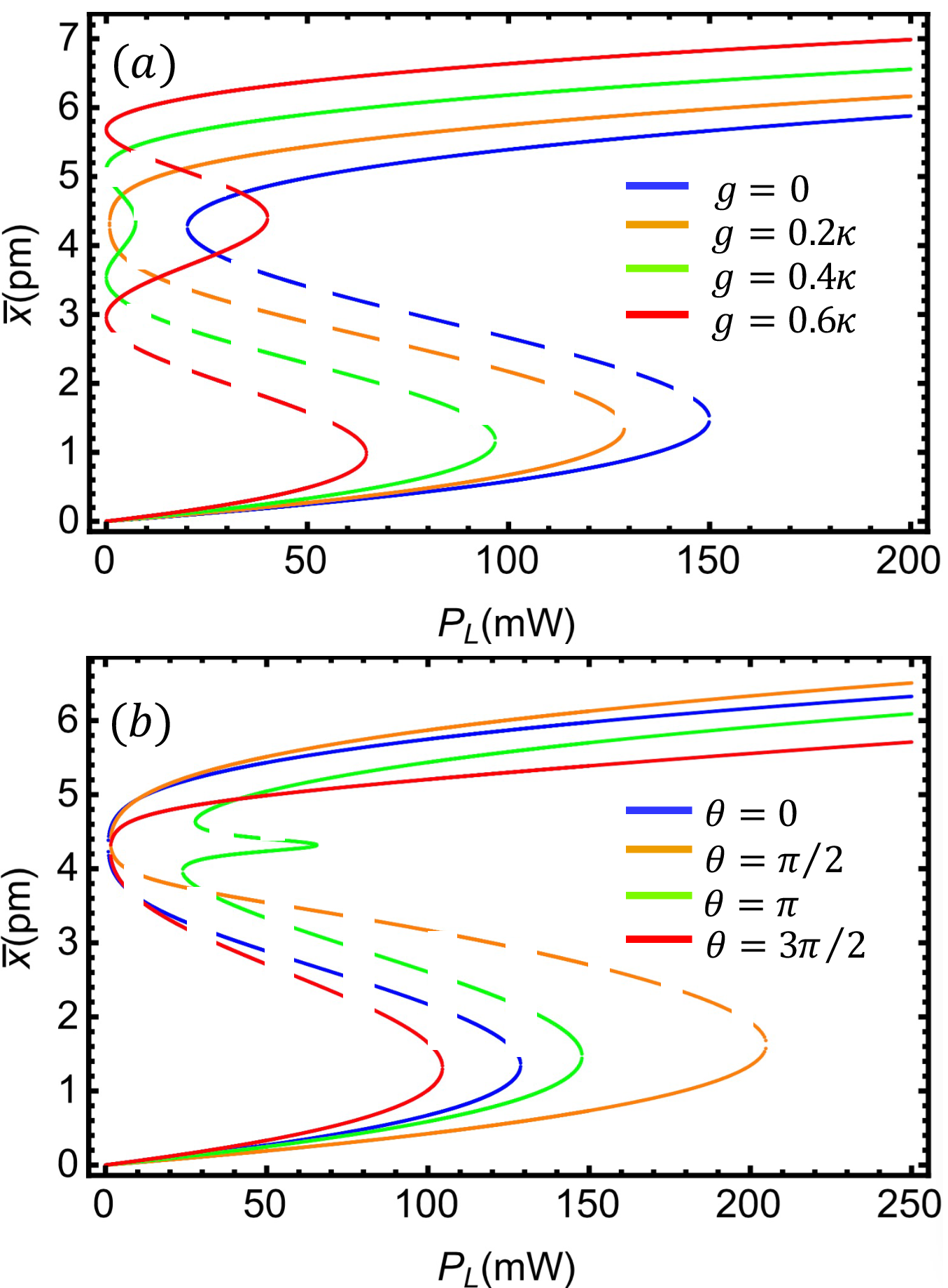}
\end{center}
\caption{Mean displacement of the movable mirror as a function of the laser power $P_L$ for (a) different values of $g$ with $\theta=0$; and (b) different values of phase $\theta$ with $g=0.2\kappa$. The other parameters are $m=20~\mathrm{ng}$, $G/2\pi=-12~ \mathrm{GHz/nm}$, $\Gamma_m/2\pi=41~\mathrm{kHz}$, $\kappa/2\pi=15~ \mathrm{MHz}$, $\Omega_m/2\pi=51.8~\mathrm{MHz}$, and $\Delta=-\Omega_m$.}
\label{fig:2}
\end{figure}
where $\Delta^\prime=\Delta-G\bar{x}$ is the effective detuning of the cavity which includes the radiation pressure. Note that the intra-cavity photon number $|\bar{a}|^2$ and the displacement of mechanical oscillator $\bar{x}$ exhibit strong dependence on the magnitude of nonlinear gain $g$ and the phase $\theta$ of the OPA. Namely, the bistability of the system will be affected by the parameters of the OPA. Figure 2 shows the displacement $\bar{x}$ varies with the power of the control field by solving Eqs.~(\ref{eq8}) numerically. The parameters $m=20~\mathrm{ng}$, $G/2\pi=-12~ \mathrm{GHz/nm}$, $\Gamma_m/2\pi=41~\mathrm{kHz}$, $\kappa/2\pi=15~ \mathrm{MHz}$, $\Omega_m/2\pi=51.8~\mathrm{MHz}$, and $\Delta=-\Omega_m$ are taken from a recent experiment~\cite{10}. The wavelength of the control field is chosen to be $532~\mathrm{nm}$. Figure 2 (a) shows the displacement $\bar{x}$ as a function of laser power $P_L$ with $\theta=0$ for $g=0,~0.2\kappa,~0.4\kappa,~0.6\kappa$, respectively. It can be seen that the mean displacement exhibits the standard S-shaped bistability when $g = 0$. As the nonlinear gain $g$ increases, the system shows a transition from bistability to tristability. Moreover, when $g$ increases, a lower threshold value of the laser power $P_L$ is needed to observe the bistable behavior and the bistable region is decreased. Next, we discuss the effect of the phase $\theta$ on the bistable behavior. Figure 2 (b) shows the hysteresis loop for the mean displacement of the mechanical oscillator with $g=0.2\kappa$ for different phases. In particular, when $\theta=\pi$, the system shows tristability behavior.

Now we consider the perturbation made by the probe field. The quantum Langevin equations for the fluctuations are given by
\begin{subequations}\label{eq9}
\begin{align}
\nonumber
&\dot{\delta a}=\left(-\frac{\kappa}{2}+i\Delta^\prime\right)\delta a-iG(\bar{a}\delta x+\delta x\delta a)\\
&+2ge^{i\theta}\delta a^*+\sqrt{\eta_c\kappa}(\epsilon_1e^{-i\delta_1 t}+\epsilon_2e^{-i\delta_2 t}),\\
&\hat{\Psi}\delta x=-\hbar G(\bar{a}^*\delta a+\bar{a}\delta a^*+\delta a^*\delta a),
\end{align}
\end{subequations}
where $\hat{\Psi}=m\left(\frac{d}{dt^2}+\Gamma_m\frac{d}{dt}+\Omega_m^2\right)$. By neglecting the nonlinear terms $-iG\delta x\delta a$ and $-\hbar G\delta a^* \delta a$, these equations of motion can be solved analytically with the linearized ansatz $\delta a=a_1^+ e^{-i\delta_1 t}+a_1^- e^{i\delta_1 t}+a_2^+ e^{-i\delta_2 t}+a_2^- e^{i\delta_2 t}$, $\delta x=x_1 e^{-i\delta_1 t}+x_1^* e^{i\delta_1 t}+x_2 e^{-i\delta_2 t}+x_2^* e^{i\delta_2 t}$, where second- and higher-order nonlinear terms are ignored. While such 
linearized dynamics can explain many phenomena arise in cavity optomechanics, the nonlinear terms $-iG\delta x\delta a$ and $-\hbar G\delta a^* \delta a$ must be taken into account for  the discussion of sum sideband generation, which is out of the frequency space of linearized dynamics. To calculate the amplitudes of the sum sidebands, we assume that the fluctuation terms $\delta a$ and $\delta x$ have the following forms~\cite{10,37}:
\begin{subequations}\label{eq10}
\begin{align}
\nonumber
\delta a=&a_1^+ e^{-i\delta_1 t}+a_1^- e^{i\delta_1 t}+a_2^+ e^{-i\delta_2 t}+a_2^- e^{i\delta_2 t}\\
&+a_s^+ e^{-i\Omega_+ t}+a_s^- e^{i\Omega_+ t}\\
\nonumber
\delta x=&x_1 e^{-i\delta_1 t}+x_1^* e^{i\delta_1 t}+x_2 e^{-i\delta_2 t}+x_2^* e^{i\delta_2 t}\\
&+x_s e^{-i\Omega_+ t}+x_s^* e^{i\Omega_+ t}
\end{align}
\end{subequations}
where $\Omega_+=\delta_1+\delta_2$. Here we only focus on the first order sideband  and sum sideband process, and thus the higher order sidebands in Eqs.~(\ref{eq10})  are ignored. Substituting Eq.~(\ref{eq10}) into Eq.~(\ref{eq9}), we obtain nine algebra equations which can be divided into two groups. The first group describes the linear response of the probe field,
\begin{subequations}\label{eq11}
\begin{align}
&\alpha(-\delta_1)a_1^+=-iG\bar{a}x_1+2ge^{i\theta}a_1^{-*}+\sqrt{\eta_c\kappa}\epsilon_1\\
&\alpha(\delta_1)a_1^-=-iG\bar{a}x_1^*+2ge^{i\theta}a_1^{+*}\\
&\sigma(\delta_1)x_1=-\hbar G(\bar{a}^* a_1^+ +\bar{a}a_1^{-*})\\
&\alpha(-\delta_2)a_2^+=-iG\bar{a}x_2+2ge^{i\theta}a_2^{-*}+\sqrt{\eta_c\kappa}\epsilon_2\\
&\alpha(\delta_2)a_2^-=-iG\bar{a}x_2^*+2ge^{i\theta}a_2^{+*}\\
&\sigma(\delta_2)x_2=-\hbar G(\bar{a}^* a_2^+ +\bar{a}a_2^{-*})
\end{align}
\end{subequations}
while the second group corresponds to the sum sideband process,
\begin{subequations}\label{eq12}
\begin{align}
&\alpha(-\Omega_+)a_s^+=-iG\bar{a}x_s+2ge^{i\theta}a_s^{-*}-iG(a_2^+ x_1+a_1^+ x_2)\\
&\alpha(\Omega_+)a_s^-=-iG\bar{a}x_s^*+2ge^{i\theta}a_s^{+*}-iG(a_2^-x_1^*+a_1^-x_2^*)\\
&\sigma(\Omega_+)x_s=-\hbar G(\bar{a}^* a_s^+ +\bar{a}a_s^{-*}+a_1^+ a_2^{-*}+a_1^{-*}a_2^+)
\end{align}
\end{subequations}
where $\alpha(y)=\kappa/2-i\Delta^\prime+i y$, $\sigma(y)=m(\Omega_m^2-i\Gamma_m y-y^2)$, $\beta=i\hbar G^2|\bar{a}|^2=-iGm\Omega_m^2\bar{x}$, $\tau(y)=\sigma(y)+\beta/\alpha(y)^*$. The solution to these Eqs.~(\ref{eq11}), (\ref{eq12}) can be obtained as follows:
\begin{figure*}[t]
\begin{center}
\includegraphics[width=0.7\textwidth]{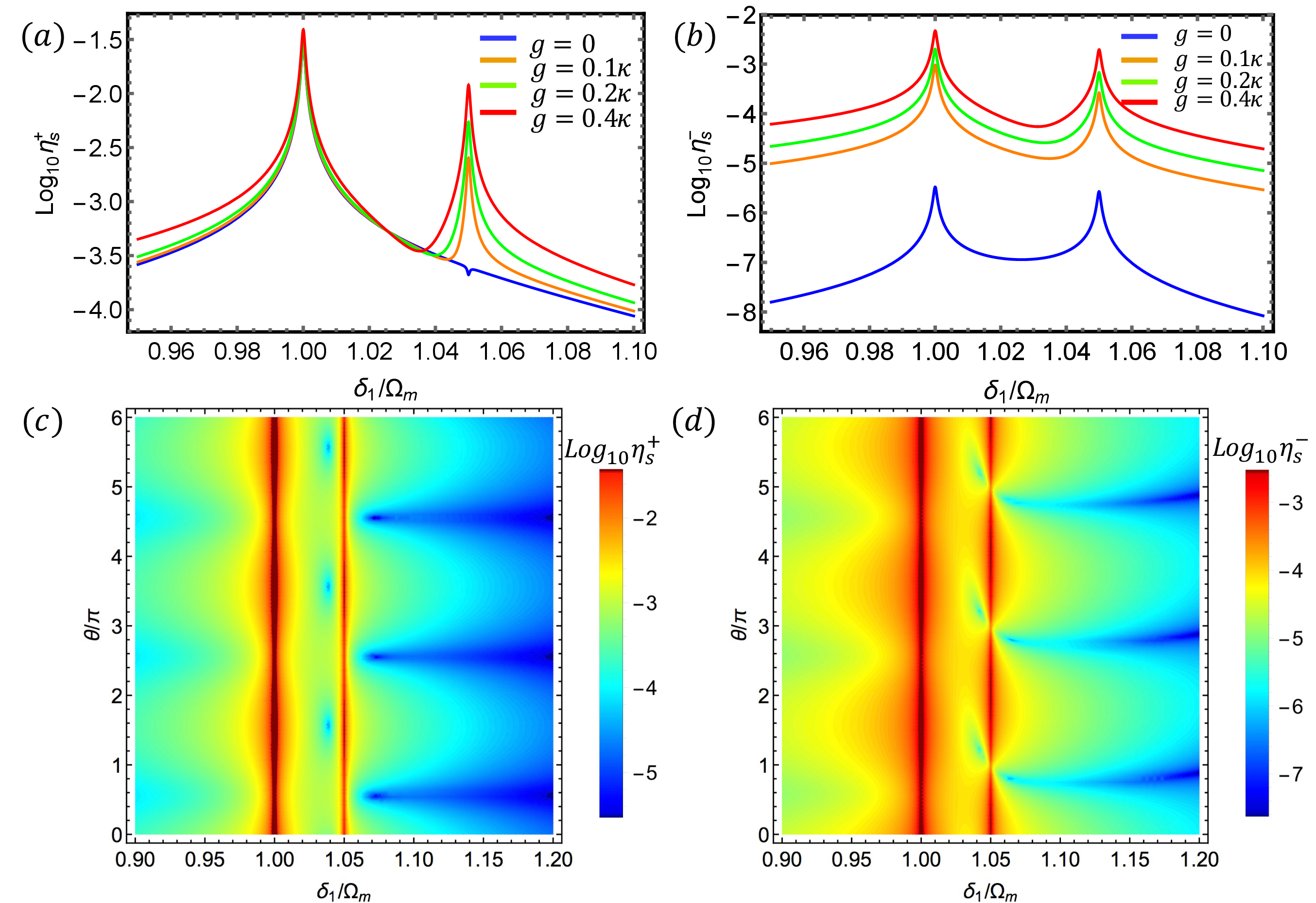}
\end{center}
\caption{Efficiencies (in logarithmic form) of (a) upper sum sideband generation and (b) lower sum sideband generation versus the frequency of the first probe field $\delta_1$ with $P_L=20 \mathrm{\mu W}$, $P_1=P_2=1 \mathrm{\mu W}$,  $\delta_2=-0.05\Omega_m$, $\theta=0$. The calculation results of (c) $\mathrm{Log}_{10}\eta_s^+$ and (d) $\mathrm{Log}_{10}\eta_s^-$ vary with $\delta_1$ and $\theta$ with $g=0.3\kappa$. The other parameters are the same as Fig.~\ref{fig:2}.}
\label{fig:3}
\end{figure*}
\begin{align}
\nonumber
&a_j^+=\frac{\alpha(\delta_j)^*\tau(\delta_j)}{\alpha(\delta_j)^*[\alpha(-\delta_j)\tau(\delta_j)-\beta]-\Xi(\delta_j)}\sqrt{\eta_c\kappa}\epsilon_j,\\
\nonumber
&x_j=\frac{-\hbar G(\bar{a}^*\alpha(\delta_j)^*+2ge^{-i\theta}\bar{a})}{\alpha(\delta_j)^*\tau(\delta_j)}a_j^+,\\
&a_j^-=\frac{i\hbar G^2\bar{a}^2+2ge^{i\theta}\sigma(\delta_j)}{-\hbar G[\bar{a}\alpha(\delta_j)+2ge^{i\theta}\bar{a}^*]}x_j^*, \quad (j=1,2) 
\end{align}
\begin{small} 
\begin{equation}
a_s^+=\frac{iG[\hbar G \bar{a}\alpha(\Omega_+)^*\xi_s-(a_2^+x_1+a_1^+x_2)\alpha(\Omega_+)^*\tau(\Omega_+)+\Theta]}{\alpha(\Omega_+)^*[\alpha(-\Omega_+)\tau(\Omega_+)-\beta]-\Xi(\Omega_+)},
\end{equation}
\end{small}
\begin{equation}
x_s=\frac{-\hbar G\left([\bar{a}^*\alpha(\Omega_+)^*+2g\bar{a}e^{-i\theta}]a_s^++\alpha(\Omega_+)^*\xi_s\right)}{\alpha(\Omega_+)^*\tau(\Omega_+)},
\end{equation}
\begin{equation}
a_s^-=\frac{-iG(\bar{a}x_s^*+a_1^-x_2^*+a_2^-x_1^*)+2ge^{i\theta}a_s^{+*}}{\alpha(\Omega_+)},
\end{equation}
where $\Xi(y)=4g^2\sigma(y)+2i\hbar G^2 g(\bar{a}^2e^{-i\theta}-\bar{a}^{*2}e^{i\theta})$, $\xi_s=a_1^+a_2^{-*}+a_1^{-*}a_2^++iG\bar{a}(a_1^{-*}x_2+a_2^{-*}x_1)/\alpha(\Omega_+)^*$, $\Theta=2ge^{i\theta}[\sigma(\Omega_+)(a_1^{-*}x_2+a_2^{-*}x_1)-\hbar G\bar{a}^*(a_1^+a_2^{-*}+a_1^{-*}a_2^+)]$. We can see that the amplitudes of sum sideband $a_s^{\pm}$ shows a strong dependence on the nonlinear gain $g$ and the phase $\theta$ of the OPA.

By using the standard input-output relations, i.e., $a_{\mathrm{out}}=a_{\mathrm{in}}-\sqrt{\eta_c\kappa}a$, we obtain the output fields (in a frame rotating at $\omega_L$) of this system as follows:
\begin{align}
\nonumber
&a_{\mathrm{out}}=\epsilon_L-\sqrt{\eta_c\kappa}\bar{a}+(\epsilon_1-\sqrt{\eta_c\kappa}a_1^+)e^{-i\delta_1 t}\\
\nonumber
&+(\epsilon_2-\sqrt{\eta_c\kappa}a_2^+)e^{-i\delta_2 t}-\sqrt{\eta_c\kappa}a_1^-e^{i\delta_1 t}\\
&-\sqrt{\eta_c\kappa}a_2^-e^{i\delta_2 t}-\sqrt{\eta_c\kappa}a_s^+e^{-i\Omega_+ t}-\sqrt{\eta_c\kappa}a_s^-e^{i\Omega_+ t}.
\end{align}
The term $\epsilon_L-\sqrt{\eta_c\kappa}\bar{a}$ denotes the output with pump frequency $\omega_L$, while the terms $-\sqrt{\eta_c\kappa}a_j^-e^{i\delta_j t}$ and $(\epsilon_j-\sqrt{\eta_c\kappa}a_j^+)e^{-i\delta_j t}$ ($j=1,2$) describe the Stokes and anti-Stokes fields, respectively. Moreover, the terms $-\sqrt{\eta_c\kappa}a_s^+$ and $-\sqrt{\eta_c\kappa}a_s^-$, describing the output with frequencies $\omega_L\pm\Omega_+$, are related to the upper and lower sum sidebands respectively. 

\section{results and discussion}
Here, we only consider sum sidebands, then the efficiency of the upper and lower sum sideband can be define as
$\eta_s^+=|-\sqrt{\eta_c\kappa}a_s^+/\epsilon_1|$ and $\eta_s^-=|-\sqrt{\eta_c\kappa}a_s^-/\epsilon_1|$ respectively, which are the ratio between amplitudes of the sum sideband and the first probe field, and thus dimensionless. In the previous work~\cite{37}, it is shown that the efficiencies of sum sideband generation exhibit peak structure for some specific values of $\delta_1$ and $\delta_2$. The specific values of $\delta_1$ ($\delta_2$) corresponding to these peaks are called as the matching conditions. 
\begin{figure}[t]
\begin{center}
\includegraphics[width=0.33\textwidth]{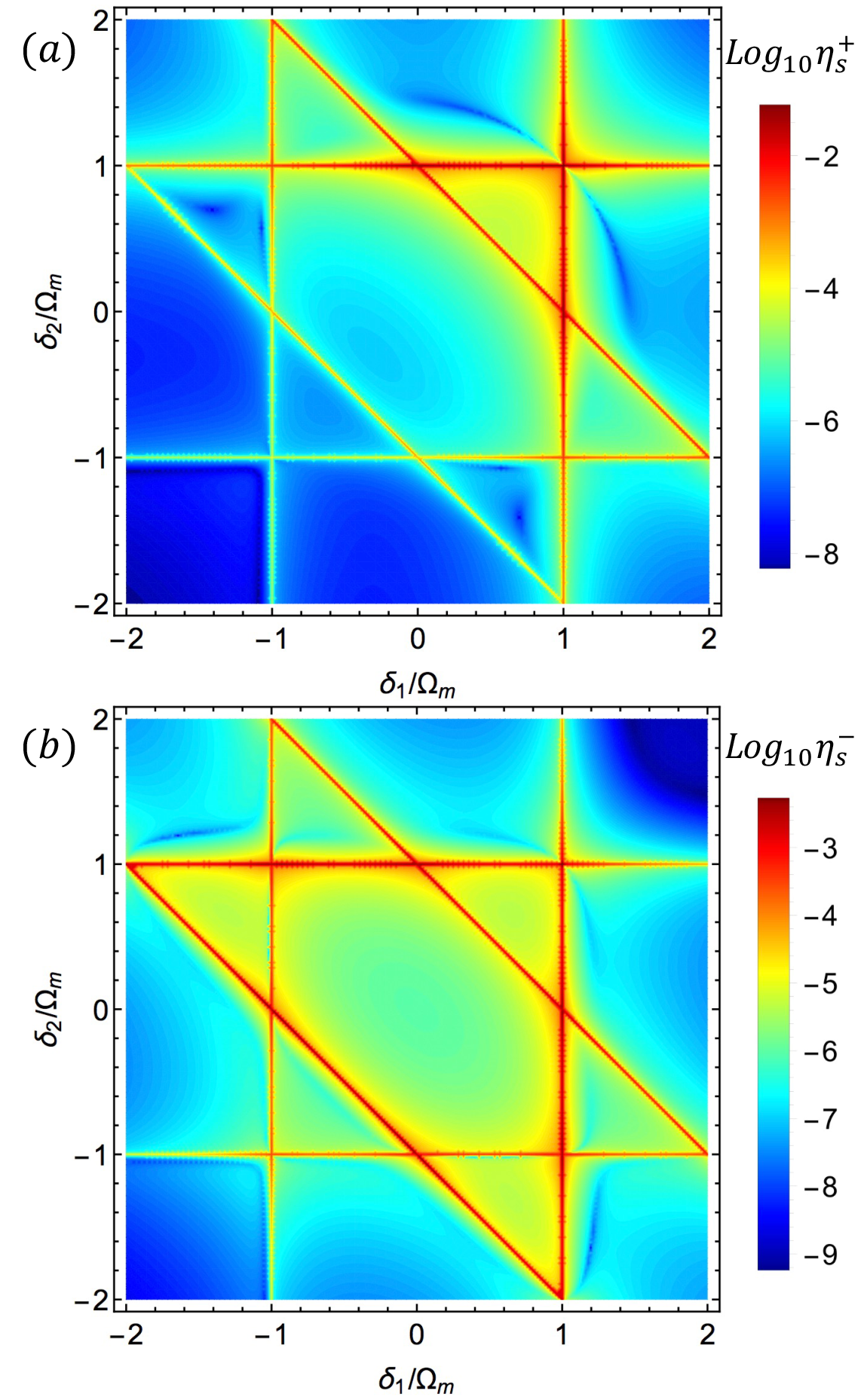}
\end{center}
\caption{Efficiencies (in logarithmic form) of (a) upper sum sideband generation and (b) lower sum sideband generation versus $\delta_1$ and $\delta_2$ with $P_L=20 \mathrm{\mu W}$, $P_1=P_2=1 \mathrm{\mu W}$, $g=0.3\kappa$, $\theta=0$. The other parameters are the same as Fig.~\ref{fig:2}.}
\label{fig:4}
\end{figure}

To illustrate the remarkable influence of the OPA on sum sideband generation, the efficiencies (in logarithmic form) of upper and lower sum sideband generation as a function of the frequency of the first probe field $\delta_1$ are shown in Fig.~\ref{fig:3}. We can clearly see that the efficiencies of sum sidebands can be significantly enhanced by the OPA. Under the weak driving field, the power of the control field $P_L=20\mu \mathrm{W}$, the probe fields $P_1=P_2=1\mu \mathrm{W}$, the frequency of the second probe field $\delta_2=-0.05\Omega_m$, and the phase of the OPA $\theta=0$. As shown in Fig.~3 (a), the efficiency (in logarithmic form) of upper sum sideband $\mathrm{Log}_{10}\eta_s^+$ has only a peak in the parameter range $0.95\Omega_m<\delta_1<1.1\Omega_m$ in the absence of the OPA, viz $g=0$. When $g\neq 0$, a new peak appears at $\delta_1=1.05\Omega_m$. Therefore, the matching condition for $\mathrm{Log}_{10}\eta_s^+$ can be modified to $\delta_1=\Omega_m$ and $\delta_1+\delta_2=\Omega_m$, which is different from the case without OPA~\cite{37}. 
Furthermore, the efficiency $\mathrm{Log}_{10}\eta_s^+$ gets larger and the linewidth widens monotonically with increasing the nonlinear gain $g$ of the OPA . To be more specific, for $g=0.4\kappa$, the efficiency $\eta_s^+$ can reach about $1\%$ at $\delta_1=1.05\Omega_m$, which is approximately 50 times larger than the case without OPA. Compared with $\eta_s^+$, the efficiency of lower sum sideband $\eta_s^-$ is enhanced more significantly in the presence of the OPA than that in a linear optomechanical system. In Fig.~\ref{fig:3}(b), We show the efficiency $\mathrm{Log}_{10}\eta_s^-$ under different strengths $g$. In the absence of the OPA, obviously, we find that the efficiency of lower sum sideband generation $\eta_s^-$ is very small (about $3\times 10^{-4} \%$ at $\delta_1=\Omega_m$) due to the weak optomechanical nonlinearity. When the OPA is considered in the optomechanical system, as expected, the efficiency of lower sum sideband generation obviously increases and the maximum values of $\eta_s^-$ is about $0.1 \%$ corresponding to the nonlinear gain strength $g=0.1\kappa$ and the detuning $\delta_1=\Omega_m$. More importantly, we find that the linewidth of the efficiency $\mathrm{Log}_{10}\eta_s^\pm$ broadens with the increasing strength $g$. The reason is that the linewidth of the OMIT window is related to the intracavity photon number~\cite{1,10}, viz $\Gamma=\Gamma_m+(2Gx_{zpf})^2|\bar{a}|^2/\kappa$, where $x_{zpf}\equiv \sqrt{\hbar/2m\Omega_m}$ is the zero-point fluctuations of the mechanical mode. From Eqs.~(\ref{eq8}), we can see that the OPA significantly increases the intracavity photon number, and thus broadens the linewidth of the efficiency $\mathrm{Log}_{10}\eta_s^\pm$. 

Next, we discuss the effect of the phase $\theta$ on the efficiencies of sum sideband generation. In Fig.~\ref{fig:3}(c) and (d), the efficiencies of upper and lower sum sideband as a function of detuning $\delta_1$ and phase $\theta$ of the OPA are plotted respectively. We can see that the efficiencies of the sum sideband are sensitive to the variation of the phase of the OPA. Specifically, when $\delta_1\in [1.05\Omega_m, 1.2\Omega_m]$, the effect of the phase $\theta$ on the efficiencies of sum sideband becomes more obviously. 

In the presence of the OPA, we can see that there is a new peak appearing in Fig.\ref{fig:3}(a), which means that there is a new matching condition for $\mathrm{Log}_{10}\eta_s^+$ achieving the maximum value. In order to see this more clearly, calculation result of efficiency (in logarithmic form) of upper sum sideband generation as functions of both $\delta_1$ and $\delta_2$ is shown in Fig.\ref{fig:4}(a), where the efficiency of upper sum sideband generation exhibits peak structure for some specific values of $\delta_1$ and $\delta_2$. From fig.~\ref{fig:4}, one can identify the matching conditions for upper and lower sum sideband generation. As is shown in previous work~\cite{37}, the upper sum sideband is enhanced when $\delta_1\rightarrow\pm\Omega_m$ and $\delta_2\rightarrow\pm\Omega_m$. By considering the effect of OPA, there are two new matching conditons  for upper sum sideband generation achieving the maximum, namely, $\delta_1+\delta_2=\pm\Omega_m$. Furthermore, the efficiency of upper sum sideband generation is enhanced more significantly when $\delta_1+\delta_2=\Omega_m$ than the case of $\delta_1+\delta_2=-\Omega_m$. For the case of lower sum-sideband generation in Fig. 4(b), the matching conditions are the same as that of upper sum sideband. The peak values of the lower sum sideband are enhanced for several orders of magnitude under the influence of the OPA.

\begin{figure}
\begin{center}
\includegraphics[width=0.33\textwidth]{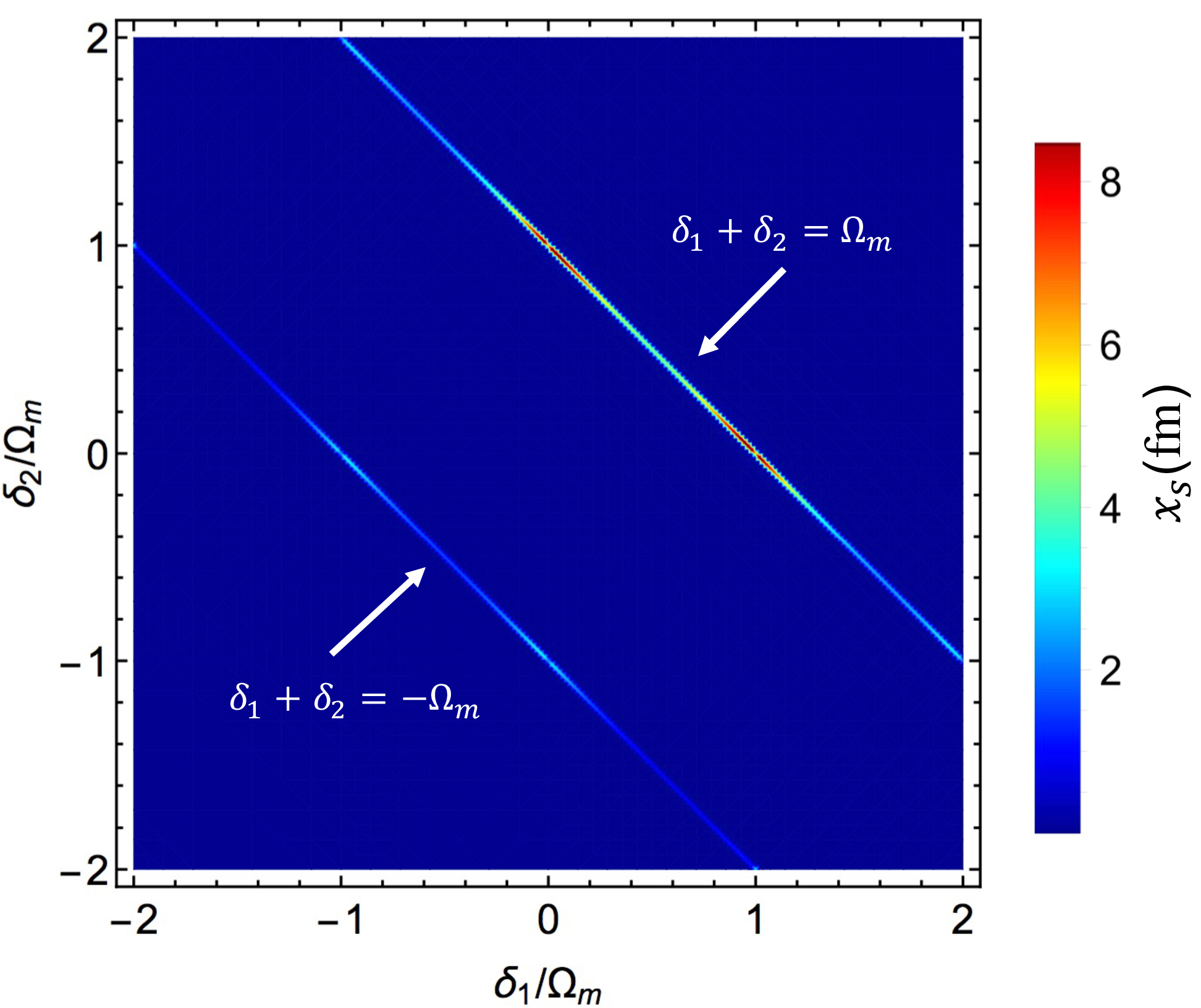}
\end{center}
\caption{The amplitude of the mechanical oscillation at the sum sideband in unit of femtometer varies with $\delta_1$ and $\delta_2$. The parameters are the same as Fig.~\ref{fig:4}.}
\label{fig:5}
\end{figure}

The physical interpretation of the new matching condition for sum sideband generation relies on the features of the mechanical oscillation at the sum sideband. Fig.~\ref{fig:5} plots the amplitude of the mechanical oscillation at the sum sideband as a function of the detuning $\delta_1$ and $\delta_2$. It is shown that $x_s$ becomes remarkable on the lines $\delta_1+\delta_2=\pm\Omega_m$, which are slightly different from the case without OPA. In the absence of OPA~\cite{37}, $x_s$ can be reinforced on the line $\delta_1+\delta_2=\Omega_m$ and at four points $(\delta_1,\delta_2)=(0,\pm\Omega_m)$ and $(\pm\Omega_m, 0)$. Furthermore, the amplitude of the mechanical oscillation at the sum sideband is enhanced more significantly on the line $\delta_1+\delta_2=\Omega_m$ than the case of $\delta_1+\delta_2=-\Omega_m$, especially in the region around points $(\delta_1,\delta_2)=(0,\Omega_m)$ and $(\Omega_m,0)$.
\section{CONCLUSION}
In conclusion, we investigate the effect of OPA on the sum sideband in an optomechanical system with double probe fields driven. In such a resonator, increasing gain of the OPA can considerably enhance the coupling between the movable mirror and the cavity field. It is shown that the sum sideband can be largely enhanced in the present of the OPA. Here we find a new matching condition for the upper sum sideband. Moreover, the lower sum sideband can be enhanced with several orders of magnitude by tuning the nonlinear gain and the phase of the OPA. These results are helpful to better understand the propagation of light in nonlinear optomechanical devices and provides potential applications to the precision measurement and optical communications.

\section*{Acknowledgments}
 This work was supported by the
National Natural Science Foundation of the People's Republic of
China (NSFC) (No. 61604045).

%\begin{thebibliography}{000} %for 3 digits
%\begin{thebibliography}{00}  %for 2 digits


\begin{thebibliography}{00}    %for 1 digit
\begin{spacing}{1}
\bibitem{1}M. Aspelmeyer, T. J. Kippenberg, and F. Marquardt, Cavity optomechanics, Rev. Mod. Phys. {\bf 86}, 1391 (2014).
\bibitem{2}M. Aspelmeyer, P. Meystre, and K. Schwab, Quantum optomechanics, Phys.  Today {\bf 65}, 29 (2012).
\bibitem{3}T. J. Kippenberg and K. J. Vahala, Cavity optomechanics: Backaction at the mesoscale, Science {\bf 321}, 1172 (2008).
\bibitem{4}C. K. Law, Interaction between a moving mirror and radiation pressure: A Hamiltonian formulation, Phys. Rev. A {\bf 51}, 2537 (1995).
\bibitem{5}C. Genes, D. Vitali, and P. Tombesi, Emergence of atom-light-mirror entanglement inside an optical cavity, Phys. Rev. A {\bf 77}, 050307(R) (2008).
\bibitem{6}A. Farace and V. Giovannetti, Enhancing quantum effects via periodic modulations in optomechanical systems, Phys. Rev. A {\bf 86}, 013820 (2012).
\bibitem{7}H. Shi and M. Bhattacharya, Quantum mechanical study of a generic quadratically coupled optomechanical system, Phys. Rev. A {\bf 87}, 043829 (2013).
\bibitem{8}J.-Q. Liao, C. K. Law, L.-M. Kuang, and F. Nori, Enhancement of mechanical effects of single photons in modulated two-mode optomechanics, Phys. Rev. A {\bf 92}, 013822 (2015).
\bibitem{9}W. Qin, A. Miranowicz, P.-B. Li, X.-Y. L\"{u}, J. Q. You, and F. Nori, Exponentially Enhanced Light-Matter Interaction, Cooperativities, and Steady-State Entanglement Using Parametric Amplification, Phys. Rev. Lett {\bf 120}, 093601 (2018).
\bibitem{10}S. Weis, R. Rivi\`{e}re, S. Del\'{e}glise, E. Gavartin, O. Arcizet, A. Schliesser, T. J. Kippenberg, Optomechanically Induced Transparency, Science {\bf 330}, 1520 (2010).
\bibitem{11}S. Gr\"{o}blacher, K. Hammerer, M. R. Vanner, and M. Aspelmeyer, Observation of strong coupling between a micromechanical resonator and an optical cavity field, Nature {\bf 460}, 724 (2009).
\bibitem{12}E. Verhagen, S. Del\'{e}glise, S. Weis, A. Schliesser, and T. J. Kippenberg, Quantum-coherent coupling of a mechanical oscillator to an optical cavity mode, Nature {\bf 482}, 63 (2012).
\bibitem{13}C. F. Ockeloen-Korppi, E. Damsk\"{a}gg, J.-M. Pirkkalainen, M. Asjad, A. A. Clerk, F. Massel, M. J. Woolley, and M. A. Sillanp\"{a}\"{a}, Stabilized entanglement of massive mechanical oscillators, Nature {\bf 556}, 478 (2018).
\bibitem{14}C. M. Caves, Quantum-Mechanical Radiation-Pressure Fluctuations in an Interferometer, Phys. Rev. Lett. {\bf 45}, 75 (1980).
\bibitem{15}A. Abramovici, W. E. Althouse, R. W. P. Drever, Y. Gürsel, S. Kawamura, F. J. Raab, D. Shoemaker, L. Sievers, R. E. Spero, K. S. Thorne, R. E. Vogt, R. Weiss, S. E. Whitcomb, and M. E. Zucker, LIGO: The laser interferometer gravitational-wave observatory, Science {\bf 256}, 325 (1992).
\bibitem{16}V. Braginsky, and S. P. Vyatchanin, Low quantum noise tranquilizer for Fabry-Perot interferometer, Phys. Lett. A {\bf 293}, 228 (2002).
\bibitem{17}A. Schliesser, O. Arcizet, R. Rivi\`{e}re, G. Anetsberger, and T. J. Kippenberg, Resolved-sideband cooling and position measurement of a micromechanical oscillator close to the Heisenberg uncertainty limit, Nature Physics {\bf 5}, 509 (2009).
\bibitem{18}P. Verlot, A. Tavernarakis, T. Briant, P.-F. Cohadon, and A. Heidmann, Backaction Amplification and Quantum Limits in Optomechanical Measurements, Phys. Rev. Lett. {\bf 104}, 133602 (2010).
\bibitem{19}A. Schliesser, R.Rivi\`{e}re, G. Anetsberger, O. Arcizet, and T. J. Kippenberg, Resolved-sideband cooling of a micromechanical oscillator, Nature physics {\bf 4}, 415 (2008).
\bibitem{20}C. Genes, D. Vitali, P. Tombesi, S. Gigan, and M. Aspelmeyer, Ground-state cooling of a micromechanical oscillator: Comparing cold damping and cavity-assisted cooling schemes, Phys. Rev. A {\bf 77}, 033804 (2008).
\bibitem{21}X. Chen, Y.-C. Liu, P. Peng, Y. Zhi, and Y.-F. Xiao, Cooling of macroscopic mechanical resonators in hybrid atom-optomechanical systems, Phys. Rev. A {\bf 92}, 033841 (2015).
\bibitem{22}W. Nie, A. Chen, and Y. Lan, Cooling mechanical motion via vacuum effect of an ensemble of quantum emitters, Opt. Express {\bf 23}, 30970 (2015).
\bibitem{23}D.-G. Lai, F. Zou, B.-P. Hou, Y.-F. Xiao, and J.-Q. Liao, Simultaneous cooling of coupled mechanical resonators in cavity optomechanics, Phys. Rev. A {\bf 98}, 023860 (2018).
\bibitem{24}G. S. Agarwal and S. Huang, Electromagnetically induced transparency in mechanical effects of light, Phys. Rev. A {\bf 81}, 041803(R) (2010).
\bibitem{25}A. H. Safavi-Naeini, T. P. Mayer Alegre, J. Chan, M. Eichenfield, M. Winger, Q. Lin, J. T. Hill, D. E. Chang, and O. Painter, Electromagnetically induced transparency and slow light with optomechanics, Nature {\bf 472}, 69 (2011).
\bibitem{26}S. Huang and G. S. Agarwal, Electromagnetically induced transparency with quantized fields in optocavity mechanics, Phys. Rev. A {\bf 83}, 043826 (2011).
\bibitem{27}H. Xiong, L.-G. Si, A.-S. Zheng, X. Yang, and Y Wu, Higher-order sidebands in optomechanically induced transparency, Phys. Rev. A {\bf 86}, 013815 (2012).
\bibitem{28}S. Liu, W.-X, Yang, T. Shui, Z. Zhu, and A.-X. Chen, Tunable two-phonon higher-order sideband amplification in a quadratically coupled optomechanical system, Sci. Rep {\bf 7}, 17637 (2017).
\bibitem{29}Y. Jiao, H. L\"{u}, J. Qian, Y. Li, and H. Jing, Nonlinear optomechanics with gain and loss: amplifying higher-order sideband and group delay. New J. Phys. {\bf 18}, 083034 (2016).
\bibitem{30}H. Suzuki, E. Brown, and R. Sterling, Nonlinear dynamics of an optomechanical system with a coherent mechanical pump: Second-order sideband generation, Phys. Rev. A {\bf 92}, 033823 (2015).
\bibitem{31}M.-A. Lemonde, N. Didier, and A. A. Clerk, Nonlinear Interaction Effects in a Strongly Driven Optomechanical Cavity, Phys. Rev. Lett. {\bf 111}, 053602 (2013).
\bibitem{32}A. Kronwald and F. Marquardt, Optomechanically Induced Transparency in the Nonlinear Quantum Regime, Phys. Rev. Lett. {\bf 111}, 133601 (2013).
\bibitem{33}Y.-C. Liu, Y.-F. Xiao, Y.-L. Chen, X.-C. Yu, and Q. Gong, Parametric Down-Conversion and Polariton Pair Generation in Optomechanical Systems, Phys. Rev. Lett. {\bf 111}, 083601 (2013).
\bibitem{34}Z.-X. Liu, H. Xiong, and Y. Wu, Generation and amplification of a high-order sideband induced by two-level atoms in a hybrid optomechanical system, Phys. Rev. A {\bf 97}, 013801 (2018).
\bibitem{35}Y.-F. Jiao, T.-X. Lu, and H. Jing, Optomechanical second-order sidebands and group delays in a Kerr resonator, Phys. Rev. A {\bf 97}, 013843 (2018).
\bibitem{36}H. Xiong, Y.-W. Fan, X. Yang, and Y. Wu, Radiation pressure induced difference-sideband generation beyond linearized description, Appl. Phys. Lett. {\bf 109}, 061108 (2016).
\bibitem{37}H. Xiong, L.-G. Si, X.-Y. L\"{u}, and Y. Wu, Optomechanically induced sum sideband generation, Opt. Express {\bf 24}, 5773 (2016).
\bibitem{38}E. Gavartin, P. Verlot, and T. J. Kippenberg, A hybrid on-chip optomechanical transducer for ultrasensitive force measurements, Nat. Nanotechnol. {\bf 7}, 509 (2012).
\bibitem{39}A. G. Krause, M. Winger, T. D. Blasius, Q. Lin, and O. Painter, A high- resolution microchip optomechanical accelerometer, Nat. Photonics {\bf 6}, 768 (2012).
\bibitem{40}H. Xiong, Z.-X. Liu, and Y. Wu, Highly sensitive optical sensor for precision measurement of electrical charges based on optomechanically induced difference-sideband generation, Opt. Lett. {\bf 42}, 3630 (2017).
\bibitem{41}J. D. Cohen, S. M. Meenehan, G. S. MacCabe, S. Gr\"{o}blacher, A. H. Safavi-Naeini, F. Marsili, M. D. Shaw, and O. Painter, Phonon counting and intensity interferometry of a nanomechanical resonator, Nature {\bf 520}, 522 (2015).
\bibitem{42}X.-Y. L\"{u}, Y. Wu, J. R. Johansson, H Jing, J. Zhang, and F. Nori, Squeezed Optomechanics with Phase-Matched Amplification and Dissipation, Phys. Rev. Lett. {\bf 114}, 093602 (2015).
\bibitem{43}S. Huang and G. S. Agarwal, Normal-mode splitting in a coupled system of a nanomechanical oscillator and a parametric amplifier cavity, Phys. Rev. A {\bf 80}, 033807 (2009).
\bibitem{44}S. Huang and G. S. Agarwal, Enhancement of cavity cooling of a micromechanical mirror using parametric interactions, Phys. Rev. A {\bf 79}, 013821 (2009).
\bibitem{45}G. S. Agarwal and S. Huang, Strong mechanical squeezing and its detection, Phys. Rev. A {\bf 93}, 043844 (2016).
\bibitem{46}A. Xuereb, M. Barbieri, and M. Paternostro, Multipartite optomechanical entanglement from competing nonlinearities, Phys. Rev. A {\bf 86}, 013809 (2012).
\bibitem{47}S. Huang and G. S. Agarwal, Robust force sensing for a free particle in a dissipative optomechanical system with a parametric amplifier, Phys. Rev. A {\bf 95}, 023844 (2017).
\bibitem{48}S. Pina-Otey, F. Jim\'{e}nez, P. Degenfeld-Schonburg, and C. Navarrete-Benlloch, Classical and quantum-linearized descriptions of degenerate optomechanical parametric oscillators, Phys. Rev. A {\bf 93}, 033835 (2016).
\end{spacing}
\end{thebibliography}
\end{document}